\documentstyle[prd,aps]{revtex}
\newcommand{\beq}{\begin{equation}}
\newcommand{\eeq}{\end{equation}}
\newcommand{\bea}{\begin{eqnarray}}
\newcommand{\eea}{\end{eqnarray}}

\begin{document}

\def\gamh{\Gamma_H}
\def\eb{E_{\rm beam}}
\def\deb{\Delta E_{\rm beam}}
\def\sigm{\sigma_M}
\def\sigmmax{\sigma_M^{\rm max}}
\def\sigmmin{\sigma_M^{\rm min}}
\def\sige{\sigma_E}
\def\dsigm{\Delta\sigma_M}
\def\mh{M_H}
\def\lyear{L_{\rm year}}

\def\wstar{W^\star}
\def\zstar{Z^\star}
\def\ie{{\it i.e.}}
\def\etal{{\it et al.}}
\def\eg{{\it e.g.}}
\def\pzero{P^0}
\def\mt{m_t}
\def\mpzero{M_{\pzero}}
\def\mev{~{\rm MeV}}
\def\gev{~{\rm GeV}}
\def\gam{\gamma}
\def\lsim{\mathrel{\raise.3ex\hbox{$<$\kern-.75em\lower1ex\hbox{$\sim$}}}}
\def\gsim{\mathrel{\raise.3ex\hbox{$>$\kern-.75em\lower1ex\hbox{$\sim$}}}}
\def\ntc{N_{TC}}
\def\epem{e^+e^-}
\def\tauptaum{\tau^+\tau^-}
\def\lplm{\ell^+\ell^-}
\def\anti{\overline}
\def\mz{M_Z}
\def\mw{M_W}
\def\fbi{~{\rm fb}^{-1}}
\def\mupmum{\mu^+\mu^-}
\def\rts{\sqrt s}
\def\sigrts{\sigma_{\tiny\rts}^{}}
\def\sigrtssq{\sigma_{\tiny\rts}^2}
\def\sigrtsprime{\sigma_{E}}
\def\nsigrts{n_{\sigrts}}
\def\gampzero{\Gamma_{\pzero}}
\def\pzerop{P^{0\,\prime}}
\def\mpzerop{M_{\pzerop}}

\draft
\input epsf

\twocolumn[\hsize\textwidth\columnwidth\hsize\csname
@twocolumnfalse\endcsname

\title{A New Technique for Determining the Properties
of a Narrow $s$-channel Resonance at a Muon Collider}
\author{R. Casalbuoni$^{a,b}$, A. Deandrea$^c$, S. De Curtis$^b$,
\\ D. Dominici$^{a,b}$, R. Gatto$^d$ and J. F. Gunion$^e$}
\address{\phantom{ll}}
\address{\it{$^a$Dipartimento di Fisica, Universit\`a di Firenze, I-50125
Firenze, Italia
\\
$^b$I.N.F.N., Sezione di Firenze, I-50125 Firenze, Italia\\
$^c$Institut Theor. Physik,  Heidelberg University, D-69120
             Heidelberg, Germany
\\ $^d$D\'epart. de Physique Th\'eorique, Universit\'e de
Gen\`eve, CH-1211 Gen\`eve 4, Suisse
\\
$^e$Department of Physics, University of California,
Davis, CA 95616, USA}}
\date{\today}
\maketitle
\begin{abstract}
We explore an alternative to the usual procedure of scanning
for determining the properties of a narrow $s$-channel resonance.
By varying the beam energy resolution while sitting on the resonance peak,
the width and branching ratios of the resonance can be determined.
The statistical accuracy achieved 
is superior to that of the usual scan procedure in the case of
a light SM-like Higgs boson with $\mh>130\gev$ or for
the lightest pseudogoldstone boson of 
a strong electroweak breaking model if $\mpzero>150\gev$.
\end{abstract}
\pacs{PACS: 14.80.Bn, 14.80Mz, 12.20.Fv   \hskip 5.5 cm\\
UCD-99-7~~  UGVA-DPT-1999 04-1035 \hskip 1.5 cm
hep-ph/9904253}
\vskip2pc]

The optimal means for studying $s$-channel resonance production
at a lepton collider 
depends critically on the resonance width $\Gamma$ compared to
the natural resolution, $\sigma_E$ (the Gaussian width), 
in $E\equiv\sqrt s$. If $\Gamma$
is at least as large as the natural value of $\sigma_E$ at $E=M$, the best
procedure is to simply scan the resonance using measurements at $E=M$
and at several locations off the resonance peak. However,
either a light Standard Model (SM) (or SM-like) Higgs boson, $H$, 
or the lightest pseudogoldstone boson of a strong electroweak
breaking model, $\pzero$, will typically have 
such a small width that $\Gamma\gsim \sigm$ can only be
achieved by compressing the beams to $R$ values far smaller than
the natural value, which can only be accomplished with substantial
loss of instantaneous luminosity.
Writing $\deb/\eb=0.01R$, with $R$ in percent,
$\sige={0.01\,R(\%)E\over\sqrt 2}\sim 2\mev
\left({E\over 100\gev}\right)\left({R\over 0.003\%}\right)\,,$
where $R=0.003\%$ is the best resolution that can be achieved at a 
$\rts \sim 100\gev$ muon collider. For comparison, $\gamh
\sim 1,2,4,16\mev$ at $\mh=50,100,130,150\gev$ 
while $\gampzero\sim 2,5,11,21\mev$ at $\mpzero=50,100,150,200\gev$
(for $\ntc=4$ and the model parameter choices of Ref.~\cite{mumu}).

At a muon collider, the natural $R$ range yielding maximal luminosity is
$R=0.1\%\div0.15\%$.
${\cal L}$ declines rapidly as $R$ is decreased below this range.
For $E=100\gev$, one finds \cite{reportusa}
(see also Table 5 in \cite{reporteurope})
$\lyear=1.2\fbi\,\left({R\over 0.12\%}\right)^{0.67362}$
for $0.003\%\lsim R\lsim 0.12\%$,
yielding $\lyear=0.1\fbi$ ($0.22\fbi$) at $R=0.003\%$ ($0.01\%$).
We will presume that a machine specifically designed for operation at 
any given energy within a factor of 2 of $E=100\gev$ will
have the same $\lyear$. Variations of the luminosity of a machine
designed for operation at $E=100\gev$, but run at some other energy
will not be accounted for.

A scan determination of the properties
of a SM-like Higgs boson at a muon collider
has been studied in Ref.~\cite{bbgh}
(see also \cite{gunmumu}), where it was shown that
the accuracy of the $R=0.003\%$ measurements might make it possible
to distinguish a SM Higgs from the lightest Higgs of 
the minimal supersymmetric model (MSSM). 
Precision measurements of the properties of the $\pzero$
would also be possible via scanning \cite{mumu} and very valuable.
In this letter, we explore an alternative technique to scanning.
The new procedure consists of collecting two sets of data at $E=M$,
one while operating with $\Gamma>\sigm$ (or at least $\sim \sigm$) and
one with $\sigm>\Gamma$. We demonstrate that this alternative procedure
leads to smaller statistical errors for resonance properties
than the conventional scanning procedure for some ranges
of $\mh$ and $\mpzero$.

We presume that the initial scan required to precisely locate the resonance
provides a rough determination of its width.
A Breit-Wigner form for the resonance cross
section convoluted with a Gaussian energy distribution in $E$
of width $\sigm$ centered at $E=M$ yields the effective cross section
$\sigma_c$. For a given final state $F$, 
one finds (see \cite{bbgh}):
$\sigma_c^F={4\pi B_{\mupmum}B_F\over M^2}~{\rm for}~\Gamma\gg\sigm$
and
$\sigma_c^F={4\pi B_{\mupmum}B_F \over M^2}{\Gamma \sqrt\pi\over 2\sqrt
2\sigm}~{\rm for}~\Gamma\ll\sigm\,.$
Here $B_{\mupmum}$ and $B_F$ are the $\mupmum$
and $F$ branching ratios.
If we operate the collider at 
$\sigmmin\ll\Gamma$ and $\sigmmax\gg\Gamma$,
we find $\sigma_c^F(\sigmmin)/\sigma_c^F(\sigmmax)=[2\sqrt
2\sigmmax]/[\Gamma\sqrt \pi]$. Since $\sigm$ will be
precisely known \cite{reporteurope}, $\Gamma$ can
be determined from the ratio. The best determination of
$\Gamma$ is obtained by combining results for all
viable final state channels $F$. Once $\Gamma$ is known,
the two measurements of $\sigma_c^F$ 
determine $B_{\mupmum}B_F$ for any $F$.
The total width and branching ratios (converted to partial
widths using $\Gamma$)
are key to understanding the nature of the  resonance.

\begin{figure}[h]
\epsfysize=2.0in
\centerline{\epsffile{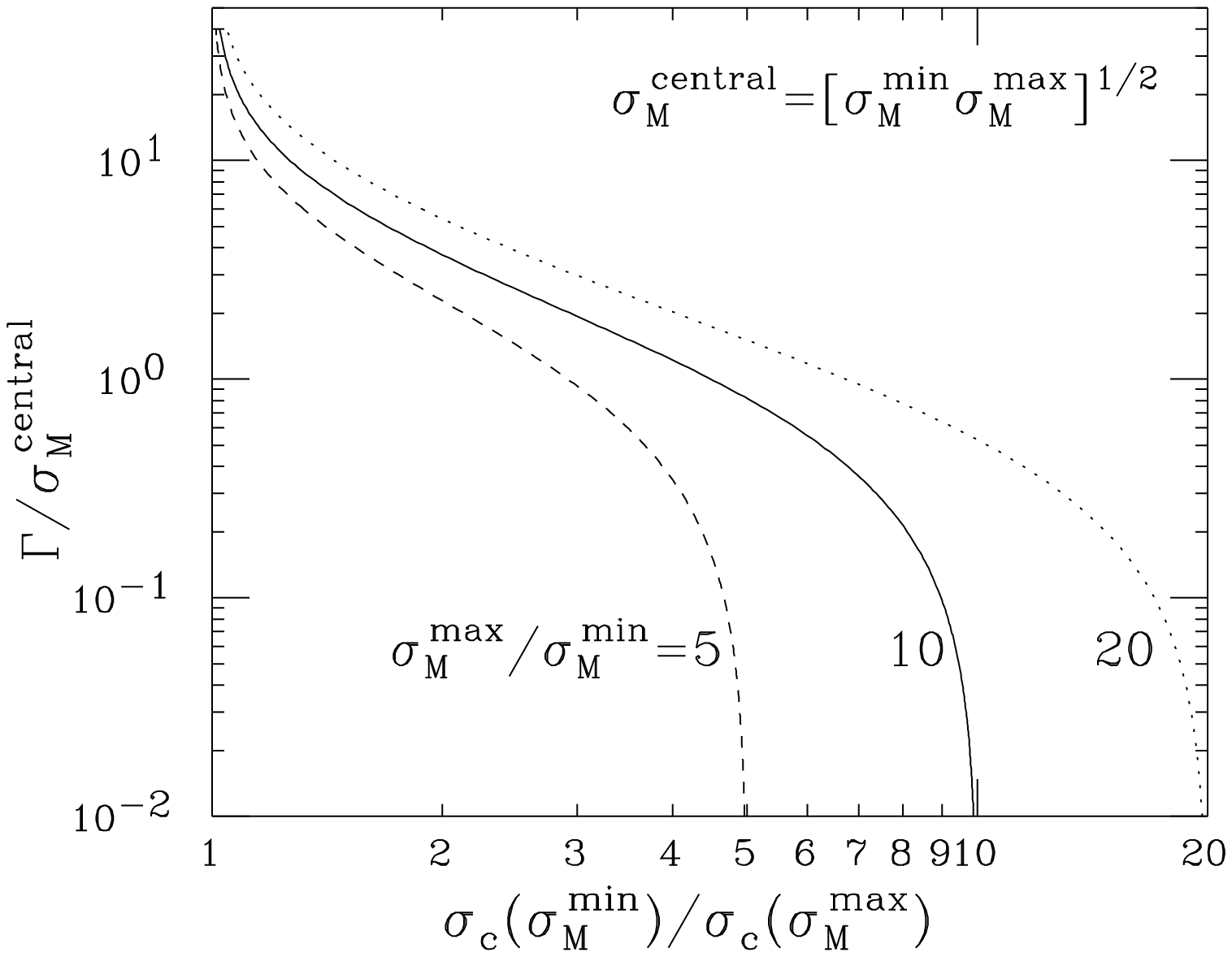}}
\smallskip
\caption{
$\Gamma/\sigm^{\rm central}$ plotted as a function of the cross section
ratio $\sigma_c(\sigmmin)/\sigma_c(\sigmmax)$
for the indicated values of $\sigmmax/\sigmmin$ keeping
$\sigm^{\rm central}\equiv\protect
\sqrt{\sigmmax\sigmmin}$ fixed.}
\label{dlgdlr}
\end{figure}

In practice, $\sigmmax/\sigmmin$ 
will be limited in size. We define 
$\sigm^{\rm central}=\sqrt{\sigmmax\sigmmin}$ and
compute $r_c\equiv\sigma_c(\sigmmin)/\sigma_c(\sigmmax)$
(we temporarily drop the final state $F$ label)
as a function of $\Gamma$. In Fig.~\ref{dlgdlr},
we plot $\Gamma/\sigm^{\rm central}$ as a function of $r_c$.  
We denote the magnitude of the slope in the log -- log plot by $|s|$.
For a known $\sigm^{\rm central}$, the $|s|$
at any $\Gamma/\sigm^{\rm central}$ gives the relation
$(\Delta\Gamma/\Gamma)=|s|(\Delta r_c/r_c)$,
where $\Delta r_c/r_c$ is computed by combining the
fractional statistical errors for $\sigma_c(\sigmmin)$
and $\sigma_c(\sigmmax)$ in quadrature.  
We observe that $\Gamma/\sigm^{\rm central}\sim 2\div 3$ gives the
smallest $|s|$ (and hence smallest statistical error),
although $|s|$ at $\Gamma/\sigm^{\rm central}\sim 1$ 
is not that much larger. 
The larger $\sigmmax/\sigmmin$, the smaller $|s|$
at any given $\Gamma/\sigm^{\rm central}$. For example, for
$\Gamma/\sigm^{\rm central}$ in the range  2 to 3 (near 1), 
$\sigmmax/\sigmmin=5,10,20$ gives $|s|\sim 2,1.55,1.3$ 
(2.5,1.8,1.44); $|s|\to 1$ for very large $\sigmmax/\sigmmin$.

The $\Delta r_c/r_c$ fractional statistical error depends 
upon how ${\cal L}$ behaves as a function of $\sigm$.
For the $H$ and the $\pzero$ it is best to  use
$\sigmmin$ corresponding to $R=0.003\%$ and 
$\sigmmax$ corresponding to $R=0.03\%$.
The variation of $\lyear$ given earlier
implies $\lyear=0.1\fbi$ ($0.47\fbi$)
for $R=0.003\%$ ($0.03\%$). If, for example, $\Gamma/\sigm^{\rm central}=1$, 
one finds $\sigma_c(\sigmmin)/\sigma_c(\sigmmax)=4.5$, implying that
the signal rate $S(\sigm)=\lyear(\sigm)\sigma_c(\sigm)$ is nearly 
the same for $\sigmmax$ as for $\sigmmin$. However,
the background rate $B$ is proportional
to $L$ and thus $B/S$ is a factor of 4.7 times larger at $\sigmmax$
than at $\sigmmin$. Consequently, the statistical error in the measurement
of $\sigma_c(\sigmmax)$ will be worse than for 
$\sigma_c(\sigmmin)$ for the same $S$.
For a given running time at a given $\sigm$,
one must compute the channel-by-channel $S$ and $B$ rates,
compute the fractional error in $\sigma_c(\sigm)$
for each channel, and then combine all channels
to get the net $\sigma_c(\sigm)$ error. 
This must be done for $\sigm=\sigmmin$ 
and $\sigm=\sigmmax$. One then computes the net 
$r_c$ and net $\sigma_c$ errors following standard procedures.
The error $\Delta\sigma_c/\sigma_c$ is minimized by running only
at $\sigmmin$, but $\Delta r_c/r_c$ is typically
minimized for $t(\sigmmin)/t(\sigmmax)\lsim 1$. For the SM Higgs, 
a good compromise is to take $t(\sigmmin)/t(\sigmmax)=1$. 

\medskip
\noindent {\bf{SM-Higgs boson}} -  
Below 110 GeV, the width of the  Higgs
increases approximately linearly with the mass (aside from logarithmic
effects due to the running of the quark masses) which means that the
ratio $\gamh/\sigm$ is approximately constant. By choosing
$R=0.003\%$ we get $\gamh/\sigm\approx 1$.
The analysis at a muon collider done in Ref. \cite{gunmumu}
gives statistical errors for a three-point scan using scan points
at $E=M,~E=M\pm 2\sigm$ and $R=0.003\%$, assuming
$L=0.4\fbi$ total accumulated luminosity
(corresponding to 4 years of operation), with $L/5$ employed
at $E=M$, $2L/5$ at $E=M+2\sigm$ and $2L/5$ at $E=M-2\sigm$. The results
of that analysis are summarized in Table~\ref{fmcerrors}.

Let us now compare to the $r_c$-ratio technique. 
We have followed the procedure outlined in the previous section.
We employ the same total of 4 years of operation
as considered for the three-point scan, but always with $E=\mh$.
We adopt the compromise choice of devoting two years
to running at $R=0.003\%$, accumulating $L=0.2\fbi$, and
a second two years to running at $R=0.03\%$, corresponding to [using
the luminosity scaling law given earlier] $L=0.94\fbi$
of accumulated luminosity. 
The resulting statistical errors are summarized in Table~\ref{fmcerrors}. 
We observe that the ratio technique becomes superior to the scan
technique for the larger $\mh$ values. This
is correlated with the fact that $\gamh/\sigmmin$ (where $\sigmmin$
is that for $R=0.003\%$) 
becomes substantially larger than 1 for such $\mh$. In
particular, for larger $\mh$,
$\gamh/\sigm^{\rm central}$ is in a range such that $|s|$
and, consequently, the error in $\gamh$ will be minimal.
Thus, the two techniques are actually quite complementary --- by employing
the best of the two procedures, a very reasonable determination of $\gamh$
and very precise determinations of the larger channel rates
will be possible for all $\mh$ below $2\mw$.

For the larger $\mh$ values such that
$\gamh/\sigm(R=0.003\%)$ is substantially above 1, one
could ask whether the scan-procedure errors could be reduced
by running at larger $R$. In fact,
the statistical errors for $\gamh$ are much poorer if a larger value
of $R$ is employed; the $R=0.003\%$ results are the best
that can be achieved despite the smaller luminosity at $R=0.003\%$
as compared to higher $R$ values. For example, the error in $\gamh$
for a given luminosity using $R=0.01\%$ 
can be read off from Fig.~13 of \cite{bbgh}. One finds that
$L(R=0.01\%)/L(R=0.003\%)=20,10,2$ is required in order that
the $\gamh$ statistical errors for $R=0.01\%$ be equal to those
for $R=0.003\%$ at $\mh=130,140,150\gev$, respectively. Existing
machine designs are such that 
$\lyear(R=0.01\%)/\lyear(R=0.003\%)=0.22\fbi/0.1\fbi=2.2$.  Thus,
increasing $R$ would not improve the scan-procedure errors 
until $\mh>150\gev$.

\medskip
\noindent {\bf{The lightest PNGB}} - The $s$-channel
production of the lightest neutral
pseudo-Nambu-Goldstone boson (PNGB) ($\pzero$), present in models of
dynamical breaking of the electroweak symmetry which have a
chiral symmetry larger than $SU(2)\times SU(2)$, has recently been
 explored \cite{mumu,mumualso}. The $\pzero$ is much lighter
than any other state in the models considered in \cite{mumu} ---
$10\gev<\mpzero<200\gev$ is expected.
The width $\gampzero$ as a function $\mpzero$ was summarized earlier.
For low (high) $\mpzero$ it is somewhat larger (smaller) 
than that of a SM Higgs boson. Very high $\mupmum\to\pzero$
$s$-channel production rates are predicted for typical $\mupmum\to\pzero$
coupling strength if one operates the $\mupmum$ collider
so as to have extremely small beam energy spread, $R=0.003\%$, for which
$\sigm<\gampzero$. Once discovered at the LHC (or Tevatron)
in the $\gam\gam$ mode,
the $\mupmum$ collider could quickly (in less
than a year) scan the mass range indicated by the previous
discovery (for the expected uncertainty in the mass
determination) and center on $\rts\simeq\mpzero$ to within
$<\sigm$. Using the optimal three-point scan \cite{mumu}
of the $\pzero$ resonance (with measurements
at $E=\mpzero$ and $E=\mpzero\pm 2\sigm$
using $R=0.003\%$) one can  determine with high
statistical precision all the $\mupmum\to\pzero\to F$ channel
rates and the total width $\gampzero$. For the particular
technicolor model parameters analysed in  \cite{mumu}, 4 years
of the pessimistic yearly  luminosity ($\lyear=0.1\fbi$)
devoted to the scan yields the results presented
in Fig.~19 of \cite{mumu}. Sample statistical errors for 
$\sigma_cB(\pzero\to {\rm all})$ and $\gampzero$ 
are given in Table~\ref{fmcerrorspgb}.

Let us now consider the $r_c$-ratio technique for the $\pzero$. We will
compare to the scan technique using the choices $R=0.003\%$ for $\sigm^{\rm
min}$ and $R=0.03\%$ for $\sigmmax$. This means $\sigm^{\rm
central}\sim 6.3\mev\,(\mpzero/100\gev)$, implying
that $\gampzero/\sigm^{\rm central}$ rises from $\sim 0.7$
at $\mpzero=50\gev$ to $\sim 1.6$ at $\mpzero=200\gev$.
This region is that for which the slope $|s|$ (see Fig.~\ref{dlgdlr}) 
is smallest. Consequently, the error in $\gampzero$
will be small if that for $r_c$ is. We follow the procedure
outlined earlier for computing $\Delta r_c/r_c$. 
We rescale the errors given in Fig.~19 of \cite{mumu}
to $L=0.2\fbi\,f$ (corresponding to $2f$ years of operation
at $R=0.003\%$), where $f$ will be chosen to minimize the error in $r_c$. 
We also compute $\Delta\sigma_c/\sigma_c$ for $L=0.94\fbi\,(2-f)$
devoted to $R=0.03\%$ running (corresponding to $4-2f$ years
of operation at this latter $R$).
The net errors, $\Delta\sigma_c/\sigma_c$ and
$\Delta\gampzero/\gampzero$,
computed after combining all final state channels, are 
given in Table~\ref{fmcerrorspgb}. For $\sigma_c$,
the $r_c$-ratio procedure statistical errors are similar to
the 4-year three-point scan statistical errors. The $r_c$-ratio 
procedure errors for
$\gampzero$ are smaller than the scan errors
for larger $\mpzero$ values where 
$\gampzero/\sigm^{\rm central}$ is significantly bigger than unity.

\medskip
\noindent{\bf Summary} -
We have compared the statistical accuracy with which 
the width and cross sections of a very narrow
resonance can be determined at a muon collider via the usual scan
procedure vs. a technique in which one sits on the resonance peak
and takes the ratio of cross sections for 
two different beam energy resolutions.
For the same total machine operation time, the ratio technique gives
smaller statistical errors than the scan technique for a SM-like Higgs
with $\mh>130\gev$ or a light pseudogoldstone boson with $\mpzero>120\gev$.
Further, systematic errors associated with uncertainty in $\sigm$ are
smaller for the ratio technique than for the scan technique.

\bigskip

\noindent{\bf Acknowledgements}
JFG is supported in part by
the U.S. Department of Energy and by the Davis Institute for High
Energy Physics.

\begin{table}[h]
\caption[fake]{\baselineskip 0pt Percentage errors ($1\sigma$) for $\gamh$
and $\sigma_cB(H\to b\anti b,W\wstar,Z\zstar)$
for $\mupmum\to H$.
For the scan procedure \cite{gunmumu} we use $R=0.003\%$ and
4-year luminosity of $L=0.4\fbi$, using $L/5$ at $E=\mh$, $2L/5$ at
$E=\mh+2\sigm$ and $2L/5$ at $E=\mh-2\sigm$.
For the $r_c$-ratio procedure, we assume $E=\mh$ and
accumulate $L=0.2\fbi$ at $R=0.003\%$
and $L=0.94\fbi$ at $R=0.03\%$, corresponding
to two years of running at each $R$. For efficiencies and cuts,
see \cite{bbgh}.}
\small
\begin{center}
\begin{tabular}{|c|c|c|c|c|c|c|}
\hline
 Quantity & \multicolumn{6}{c|}{Errors for the scan procedure} \\
\hline
\hline
 {\bf Mass (GeV)} & {\bf 100} & {\bf 110} &
  {\bf 120} & {\bf 130} & {\bf 140} & {\bf 150} \\
\hline
$\sigma_cB(b\anti b)$ & $ 4\%$ & $ 3\%$ &
$3\%$ & $5\%$ & $9\%$ & $28\%$ \\
\hline
$\sigma_cB(W\wstar)$ & $32\%$ & $ 15\%$ &
$10\%$ & $8\%$ & $7\%$ & $9\%$ \\
\hline
$\sigma_cB(Z\zstar)$ & $-$ & $ 190\%$ &
$50\%$ & $30\%$ & $26\%$ & $34\%$ \\
\hline
 $\gamh$ & $30\%$ &   $ 16\%$ &
$16\%$ & $18\%$ & $29\%$ &  $105\%$ \\
\hline
\hline
 Quantity & \multicolumn{6}{c|}{Errors for the $r_c$-ratio procedure} \\
\hline
\hline
 {\bf Mass (GeV)} & {\bf 100} & {\bf 110} &
{\bf 120} & {\bf 130} & {\bf 140} & {\bf 150} \\
\hline
$\sigma_cB(b\anti b)$ & $3.8\%$ & $2.8\%$ &
 $2.8\%$ & $4.4\%$ & $7.6\%$ & $21\%$ \\
\hline
$\sigma_cB(W\wstar)$ & $26\%$ & $12\%$ & 
$7.7\%$ & $5.7\%$ & $5.0\%$ & $5.6\%$ \\
\hline
$\sigma_cB(Z\zstar)$ & $-$ & $190\%$ &
$46\%$ & $25\%$ & $20\%$ & $22\%$ \\
\hline
 $\gamh$ & $45\%$ &   $25\%$ &
$20\%$ & $19\%$ & $17\%$ &  $18\%$ \\
\hline
\end{tabular}
\end{center}
\label{fmcerrors}
\end{table}

\begin{table}[h]
\caption[fake]{\baselineskip 0pt Fractional 
statistical errors ($1\sigma$) for 
$\sigma_cB(\pzero\to {\rm all})$ (combining
$b\bar b$, $\tauptaum$, $c\bar c$ and $gg$ tagged-channel rates)
and $\gampzero$ for $\mupmum\to\pzero$. The $R=0.003\%$ three-point scan
with total $L=0.4\fbi$ ($L/5$ at $E=\mpzero$,
$2L/5$ at $E=\mpzero+2\sigm$ and $2L/5$ at $E=\mpzero-2\sigm$) is compared
to the $r_c$-ratio technique with $E=\mpzero$ luminosities 
of $L=0.2\fbi \,f$ at $R=0.003\%$ and 
$L=0.94\fbi\,(2-f)$ at $R=0.03\%$ (corresponding
to $2f$ and $4-2f$ years of running, respectively).
$f$ (tabulated below) is chosen to minimize the error in $\gampzero$. 
Efficiencies, cuts and tagging procedures are from \cite{mumu}.}
\small
\begin{center}
\begin{tabular}{|c|c|c|c|c|c|c|}
\hline
 Quantity & \multicolumn{6}{c|}{Errors for the scan procedure} \\
\hline
\hline
 {\bf Mass} & {\bf 60} & {\bf 80} & {\bf $\mz$} & {\bf 110} 
& {\bf 150} & {\bf 200} \\
\hline
$\sigma_cB$ &  0.0029 & 0.0054 & 0.043 & 0.0093 & 0.012 & 0.018\\
\hline
 $\gampzero$ & 0.014 & 0.029 & 0.25 & 0.042 & 0.052 & 0.10 \\
\hline
\hline
 Quantity & \multicolumn{6}{c|}{Errors for the $r_c$-ratio procedure} \\
\hline
\hline
 {\bf Mass} & {\bf 60} & {\bf 80} & {\bf $\mz$} & {\bf 110} 
& {\bf 150} & {\bf 200} \\
\hline
$f$ & 0.8 & 0.7 & 0.6 & 0.8 & 0.9 & 1.0 \\
\hline
$\sigma_cB$ & 0.0029 & 0.0062 & 0.055 & 0.010 & 0.011 & 0.016 \\
\hline
 $\gampzero$ & 0.014 & 0.028 & 0.24 & 0.041 & 0.039 & 0.053 \\
\hline
\end{tabular}
\end{center}
\label{fmcerrorspgb}
\end{table}

\begin{references}
\vspace*{-1cm}
\bibitem{mumu}
R. Casalbuoni, S. De Curtis, A. Deandrea,
D. Dominici, R. Gatto and J.F. Gunion,
Proceedings of  the Workshop on Physics at the First
Muon Collider and at the Front End
   of a Muon Collider, Batavia, IL, 6-9 Nov 1997, eds. S. Geer
   and R. Raja. Amer. Inst. Phys., p. 772 (1998).; {\it ibidem},
   hep-ph/9809523 (to be published in Nucl. Phys.).
\bibitem{reportusa}
C.M. Ankenbrandt et al., [Muon Collider Collaboration], Status of Muon Collider
Research and Development and Future Plans, FERMILAB-PUB-98-179, January, 1999.
\bibitem{reporteurope}
B. Autin et al., Report of  Prospective Study of Muon Storage
Rings in Europe, to be published as a CERN yellow Report.
\bibitem{bbgh}
 V. Barger, M.S. Berger, J.F. Gunion, and T.Han, Phys. Rev. Lett.
 {\bf 75} (1995) 1462; {\it ibidem},  Phys. Rep. {\bf
 286} (1997) 1.
\bibitem{gunmumu}
J.F. Gunion, Proceedings of the 5th International Conference
on Physics Beyond the Standard Model, Balholm, Norway,
29 Apr -- 4 May 1997, eds. G. Eigen, P. Osland and B. Stugu, AIP (1997),
p. 234.
\bibitem{mumualso}
K. Lane, BUHEP-98-01, hep-ph/9801385.
E. Eichten, K. Lane and J. Womersley,
Phys. Rev. Lett. {\bf 80}  (1998) 5489.
\end{references}
\end{document}